\documentclass[preprint,english,review]{elsarticle}
\usepackage{graphicx}
\usepackage{epstopdf}
\usepackage{color}
\usepackage{lineno}

\usepackage{babel}

\newcommand{\kgd}{kg$\cdot$d~}
\newcommand{\kgdf}{kg$\cdot$d}

\begin{document}

\title{Final results of the EDELWEISS-II WIMP search
         using a 4-kg array of cryogenic germanium detectors with interleaved electrodes}


\author[]{The EDELWEISS Collaboration}

\author[irfu]{\\E.~Armengaud\corref{cor1}}
\author[ipnl]{C.~Augier}
\author[neel]{A.~Beno\^{\i}t}
\author[csnsm]{L.~Berg\'e}
\author[iek,fzk]{J.~Bl$\mbox{\"u}$mer}
\author[csnsm]{A.~Broniatowski}
\author[jinr]{V.~Brudanin}
\author[ipnl]{B.~Censier}
\author[csnsm]{G.~Chardin}
\author[csnsm]{M.~Chapellier}
\author[ipnl]{F.~Charlieux}
\author[oxford]{P. Coulter}
\author[iek]{G.A.~Cox}
\author[csnsm]{X.~Defay\fnref{pa-defay}}
\fntext[pa-defay]{Present address:
Department of Physics, University of Wisconsin, Madison, WI 53706, USA}
\author[ipnl]{M.~De~Jesus}
\author[csnsm]{Y.~Dolgorouki}
\author[csnsm,irfu]{J.~Domange}
\author[csnsm]{L.~Dumoulin}
\author[fzk]{K.~Eitel}
\author[jinr]{D.~Filosofov}
\author[irfu]{N.~Fourches}
\author[ipnl]{J.~Gascon}
\author[irfu]{G.~Gerbier}
\author[ipnl]{J.~Gironnet}
\author[irfu]{M.~Gros}
\author[oxford]{S. Henry}
\author[irfu]{S.~Herv\mbox{\'e}}
\author[ipnl]{A.~Juillard}
\author[fzk]{H.~Kluck}
\author[fzk]{V.~Kozlov}
\author[oxford]{H. Kraus}
\author[sheffield]{V.A.~Kudryavtsev}
\author[lsm]{P.~Loaiza}
\author[csnsm]{S.~Marnieros}
\author[irfu]{X-F.~Navick}
\author[irfu]{C.~Nones}
\author[csnsm]{E.~Olivieri}
\author[iramis]{P.~Pari}
\author[ipnl]{L.~Pattavina}
\author[irfu]{B.~Paul}
\author[sheffield]{M. Robinson}
\author[jinr]{S.~Rozov}
\author[ipnl]{V.~Sanglard}
\author[iek]{B.~Schmidt}
\author[ipnl]{S.~Scorza\fnref{pa-scorza}}
\fntext[pa-scorza]{Present address: 
Department of Physics, Southern Methodist University, Dallas, TX 75275, USA}
\author[jinr]{S.~Semikh}
\author[irfu]{A.S.~Torrento-Coello}
\author[ipnl]{L.~Vagneron}
\author[ipnl]{M.-A.~Verdier\fnref{pa-verdier}}
\fntext[pa-verdier]{Present address: 
Department of Physics, Queen's University, Kingston, ON, K7L 3N6, Canada}
\author[irfu]{R.J.~Walker}
\author[jinr]{E.~Yakushev}

\address[irfu]{CEA, Centre d'Etudes Saclay, IRFU, 
91191 Gif-Sur-Yvette Cedex, France}
\address[ipnl]{IPNL, Universit\'{e} de Lyon, Universit\'{e} Lyon 1, 
CNRS/IN2P3, 4 rue E. Fermi 69622 Villeurbanne cedex, France}
\address[neel]{CNRS-N\'{e}el, 25 Avenue des Martyrs, 
38042 Grenoble cedex 9, France}
\address[csnsm]{CSNSM, Universit\'e Paris-Sud, IN2P3-CNRS, bat 108, 91405 Orsay,  France}
\address[iek]{Karlsruhe Institute of Technology,
Institut f\"{u}r Experimentelle Kernphysik, Gaedestr. 1, 76128 Karlsruhe, Germany}
\address[fzk]{Karlsruhe Institute of Technology,
Institut f\"ur Kernphysik, Postfach 3640, 76021 Karlsruhe, Germany}
\address[iramis]{CEA, Centre d'Etudes Saclay, 
IRAMIS, 91191 Gif-Sur-Yvette Cedex, France}
\address[jinr]{Laboratory of Nuclear Problems, JINR, Joliot-Curie 6, 
141980 Dubna, Moscow region, Russia}
\address[lsm]{Laboratoire Souterrain de Modane, CEA-CNRS, 
1125 route de Bardonn\`eche, 73500 Modane, France}
\address[oxford]{University of Oxford, Department of Physics, Keble Road, Oxford OX1 3RH, UK}
\address[sheffield]{Department of Physics and Astronomy, University of Sheffield, Hounsfield Road, Sheffield S3 7RH, UK}

\cortext[cor1]{Corresponding author}

\begin{abstract}

The EDELWEISS-II collaboration has completed a direct search for
WIMP dark matter with an array of ten 400-g cryogenic germanium detectors
in operation at the Laboratoire Souterrain de Modane.
The combined use of thermal phonon sensors and charge collection electrodes with an interleaved 
geometry enables the efficient rejection 
of $\gamma$-induced radioactivity as well as near-surface interactions.
A total effective exposure of 384~\kgdf~has been achieved, 
mostly coming from fourteen months of continuous operation. 
Five nuclear recoil candidates are observed above 20 keV, 
while the estimated background is 3.0~events. 
The result is interpreted in terms
of limits on the cross-section of spin-independent interactions of  WIMPs and nucleons.
A cross-section of  $4.4\times10^{-8}$~pb is excluded at 90\%CL for a WIMP mass of 85~GeV.
New constraints are also set on models where the WIMP-nucleon scattering is inelastic.
\end{abstract}


\maketitle

\section{Introduction}

One of the key ingredients for the current cosmological concordance model is the 
existence of a dark, matter-like fluid ruling the dynamics of structures from the 
current galactic scales to the largest scales at early cosmic times~\cite{cosmo}. 
Weakly Interacting Massive Particles (WIMPs) are a generic class of dark 
matter particle with particularly appealing features. 
They appear in several extensions of the current Standard Model of 
particle physics, where thermal production mechanisms for such particles 
in the Big Bang naturally yield the order of magnitude for the observed cosmic abundance~\cite{wimp}.

A vast effort is currently dedicated to the direct detection of WIMPs 
from the Milky Way halo through the coherent elastic scattering on nuclei 
constituting a terrestrial detector~\cite{goodman,rev}. 
A roughly exponential nuclear recoil spectrum is expected 
with typical energies of a few tens of keV. 
Theoretical models predict a wide range of WIMP-nucleon scattering cross-sections. 
Current searches are approaching sensitivities for rates of
a few 10$^{-3}$ evts/kg/day,
corresponding to cross-sections for spin-independent interactions
at the level of a few $10^{-8}$ pb. This already probes some of the 
parameter space of minimal supersymmetric extensions of the Standard Model.
A better coverage of this parameter space would demand a
sensitivity down to cross-sections of $10^{-10}$~pb,
corresponding to ultra-low event rates of 10$^{-5}$ evts/kg/day.
This requires a severe control of radioactive backgrounds,
as well as very high detector performance reliability and background rejection over the entire
fiducial volume.

The EDELWEISS program aims at improving progressively the sensitivity to WIMP 
interactions using cryogenic germanium detectors operated in a radiopure
underground environment. 
The ten detectors, named Ge-ID in the following, are 400-g ultra-pure germanium crystals 
equipped with thermal Ge-NTD (Neutron Transmutation Doping) sensors and covered by electrodes with a
design of annular concentric and plain electrodes recently developed by 
the collaboration~\cite{brink,broni}. 
The combination of electrode signals
enables the definition of an inner fiducial volume within each crystal. 
Surface interactions are rejected with an efficiency greater than 
10$^{4}$~\cite{id}. 
Within the fiducial volume, the charge collection is such that, once combined with the thermal measurement of the total energy, a powerful discrimination is obtained between nuclear recoils and the background of electron recoils.

The EDELWEISS collaboration has recently published the first WIMP search results 
obtained with these detectors~\cite{idfirst} .
At the time, a sensitivity to WIMP-nucleon scattering cross-section of 10$^{-7}$pb 
had been achieved following a continuous six-month run with ten detectors 
representing a total mass of 4~kg. 
That limit was based on the observation of one nuclear recoil candidate with a 
recoil energy above 20 keV in a total effective exposure of 144 kg$\cdot$d. 
This suggested that a better sensitivity could be achieved by a further increase 
in exposure, 
with the potential of improving the experimental limits on WIMP-nucleon scattering 
cross-sections, and of understanding the origin of the observed event. 
In this paper we report the final results obtained by increasing the exposure 
 by more than a factor of two relative to~\cite{idfirst}.
We improve the measurement of the rejection for the gamma-ray background and 
the evaluation of other potential sources of backgrounds,
in order to better assess the expected number of background events in the WIMP search.

These results will be interpreted in terms of sensitivity to the cross-section
for WIMP-nucleon elastic scattering via spin-independent interactions, as well
as in the framework of the inelastic scattering model of Ref.~\cite{smithweiner}.
The final performance of the Ge-ID detectors will be interpreted in the context
of the development of the new generation of EDELWEISS detectors designed
for the next experimental phase involving a 40-kg array aiming at sensitivities
of 10$^{-9}$pb.

\section{Experimental setup and data sets}

The experimental setup is the same as in~\cite{idfirst} and will be described in 
detail in Ref.~\cite{edwsetup}. 
It is located in the Laboratoire Souterrain de Modane (LSM). 
The rock overburden equivalent to 4800 meters of water reduces the cosmic muon flux to 4~$\mu$/m$^{2}$/day. 
The flux of neutrons above 1~MeV is 10$^{-6}$ n/cm$^2$/s~\cite{lemrani}. 
The cryostat housing the detectors is protected from the ambient $\gamma$-rays by a 20-cm lead shield. 
This shield is itself surrounded by a 50-cm polyethylene shield, 
covered by a muon veto system with a 98\% geometric efficiency for throughgoing muons.

The detectors and their mode of operation are described in detail in Ref.~\cite{idfirst}. 
There are two different detector geometries.
Five of them have bevelled edges and an average mass of 370~g, 
while the other five are cylindrical and each has a mass of 410~g. 
Each detector is equipped with six sets of electrodes.
The cylindrical or bevelled edge of a crystal is covered with two planar electrodes defining a guard volume.
The flat surfaces are covered with electrodes forming concentric rings, 
biased at alternate potentials, and defining two sets of veto and fiducial electrodes. 
The typical voltages applied between fiducial electrodes range from 6V to 8V.
The fiducial volume is defined by requiring pulses of compatible amplitude and 
timing on the fiducial electrodes on both sides of the detector, 
and no other signals above the noise on veto and guard electrodes.

The bulk of the data presented here were recorded over a 
period of fourteen months from April 2009 to May 2010.
An additional data set was recorded with two detectors during an 
initial run performed between July and November 2008. 
The overall exposure corresponds to more than a twofold
 increase with respect to~\cite{idfirst}.
In the fourteen-month running period, data from the detectors were
collected for 85\% of the time, the rest being equally shared between
regular maintenance  operations (detector regenerations and cryogenic fluid refills) 
and unscheduled stops.
All heat sensors were operational, 
whereas five ionization read-out channels among 60 were defective 
or too noisy to be used: four guards and one veto.
However the associated electrodes could be set at the correct voltage and,
due to the redundancy of the fiducial volume cuts~\cite{idfirst}, 
the corresponding detectors could be used for WIMP search.
Among the recorded data, there were 325 live days of WIMP search, 
while 10.1 were devoted to gamma calibrations and 6.4 to neutron calibrations.
The corresponding values for  the data recorded in 2008 
are 92, 16.3 and 3.1 live days.

The data set consists of the digitized pulse shapes of all channels of the detectors
within a tower of either three or four detectors. An event is recorded each time the
heat signal on any detector in the tower crosses an online trigger level.
This level is continuously and automatically adjusted to the lowest
possible value that allows keeping the trigger rate below a fraction of a Hz.
As in the previous work, the data were analyzed offline by 
two independent analysis chains.
In both cases, the periods of 
data taking retained for analysis were selected
hour-by-hour solely on the basis of the measured baseline 
resolution of the heat and ionization signals.
The first analysis was kept identical to the one in Ref.~\cite{idfirst}, 
with the same selection and cut procedures.
The second analysis aimed more specifically at optimizing 
the ionization and heat signal resolutions during periods where
electronic and microphonic noises are more significant.
In both analyses, the resulting average FWHM baseline resolution is 
1.2 keV for the heat signals (ranging from 0.6 to 2.0~keV depending on the detector) and 0.9 keV for the fiducial ionization signals (ranging from 0.7 to 1.1~keV).
The better treatment of noisy periods in the second analysis
results in a 6\% increase of selected live periods.
This paper present the results of this second analysis and
the first analysis is used to provide systematic checks and comparisons.
The analysis procedure is not blind, since it
was not entirely defined in detail before analyzing the signal region, and 
the first six-month data are included in the final result.
However, all the calibration constants and selection cuts were 
defined on data samples that exclude the low-ionization yield events 
that will be selected as nuclear recoil candidates in the WIMP search.

\section{Detector calibration}

\subsection{Fiducial volume}

The fraction of the detector volume associated with the fiducial selection 
is measured by using the rate of
events in the photopeaks at 9.0 and 10.4 keV due to the decay of the 
cosmogenically-induced isotopes $^{65}$Zn and $^{68}$Ge~\cite{idfirst}. 
Such events are expected to be homogeneously distributed in the crystal. 
The resulting fiducial masses, obtained by multiplying the
fiducial fractions with the mass of the corresponding detectors,
take into account the efficiency of the cuts on the guard, veto 
and fiducial electrodes.
The volume fraction measurements were repeated with the increased 
statistics relative to the original 6-month run, with a total of
10$^4$ events in the fiducial volume.
The exposure-weighted average fiducial mass is 160 $\pm$ 5~g,
corresponding to the same value used in the calculation of the total exposure
in Ref.~\cite{idfirst}. 

\subsection{Nuclear recoil selection}

\begin{figure}
\begin{center} \includegraphics[width=12cm]{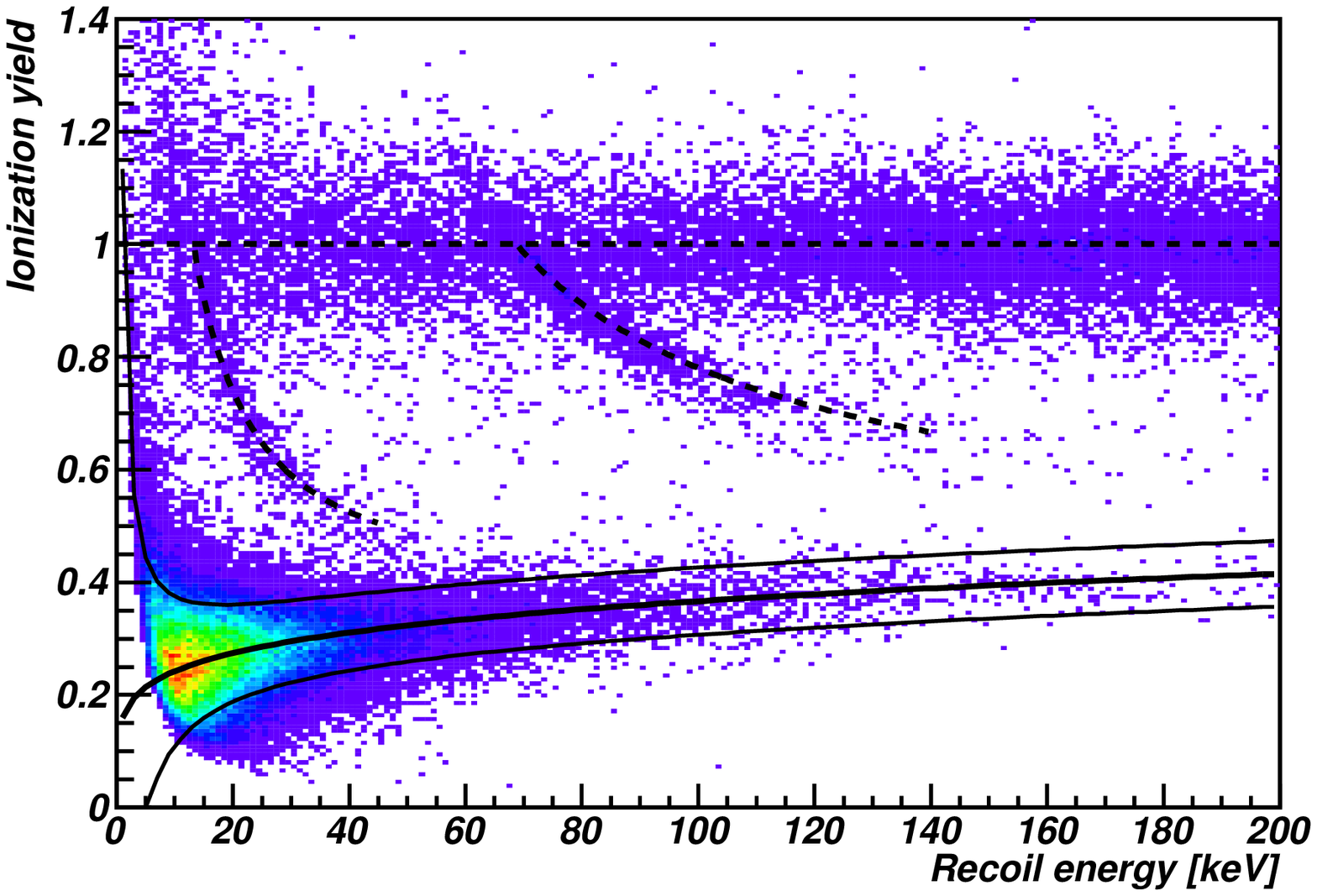} \end{center}
\caption{Distribution of the ionization yield versus recoil energy for fiducial events recorded during neutron calibrations for all Ge-ID detectors. The full lines represent the parametrization of Ref.~\cite{martineau} for nuclear recoils and the 90\%CL nuclear recoil band. In addition to pure electron and nuclear recoils, inelastic nuclear recoils are visible with associated electromagnetic energies of 13.26 and 68.75~keV, due to the desexcitation of short-lived states of $^{73}$Ge created by neutron diffusion (dashed lines).
\label{fig:neutrons}}
\end{figure}

Neutron calibrations were performed in 2008
and at the beginning and the end of the fourteen month-long run. 
The ionization yield distribution of all fiducial events recorded during these 
calibrations is shown on Fig.~\ref{fig:neutrons}. 
The distributions for nuclear recoils were measured for each detector and were
all found to be compatible with  the parametrization of Ref.~\cite{martineau}.
In particular, for each detector, the region with 90\% acceptance ($1.64 \,\sigma$) for
nuclear recoils (NR) from elastic neutron scattering,
used in the present WIMP search, is well described by the parametrization using
the measured experimental resolutions on the heat and ionization signals.
Fig.~\ref{fig:neutrons} shows the parametrization of
the average NR region.
The width of the bands varies by $\pm$15\% from detector to detector at 20 keV,
and by less than $\pm$5\% above 50 keV.

Data from neutron calibrations also enable to test the influence of thresholds 
on the nuclear recoil efficiency. 
These include the online trigger threshold, and the offline cuts applied to the heat 
and ionization energies.
In particular, Fig.~\ref{fig:neutrons} illustrates that these reduce
the acceptance for nuclear recoils at energies well below the analysis threshold of 20~keV.

\subsection{Gamma-ray rejection}

\begin{figure}
\begin{center} \includegraphics[width=11cm]{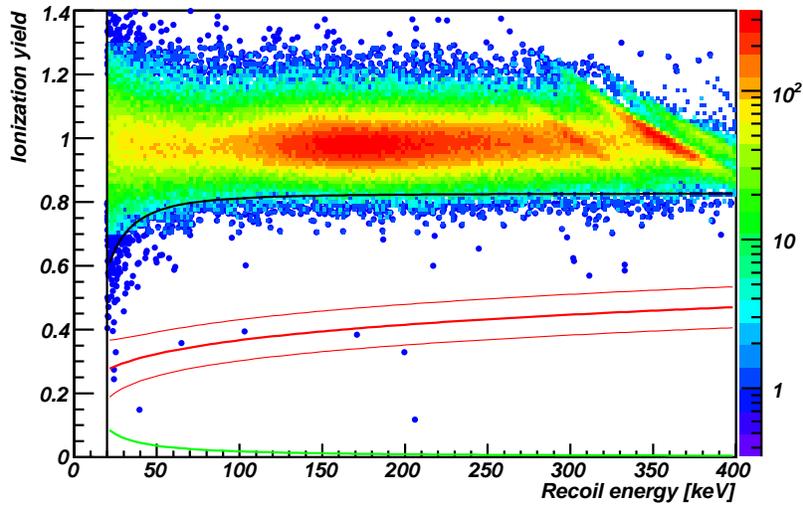} \end{center}
\caption{Distribution of the ionization yield versus recoil energy for fiducial events recorded by Ge-ID detectors during all $\gamma$-ray  calibrations regularly performed with $^{133}$Ba sources. The same period selection and quality cuts are applied than in WIMP search. The top line represents the 99.99\% lower limit of the electron recoil band for typical noise conditions. The bottom (green) line is the typical ionization threshold, while the 90\%CL nuclear recoil region is represented as a red band.
\label{fig:gamma}}
\end{figure}

The rejection factor for electron  recoils in the NR band was measured
with extensive and regular $\gamma$-ray calibrations using $^{133}$Ba sources.
The scatter plot of the measured ionization yield as a function of 
recoil energy for all calibration data is shown on  Fig.~\ref{fig:gamma}. 
Among the 3.47$\times10^{5}$ interactions above 20 keV, 
1.82$\times 10^{5}$ have recoil energies below 200~keV.
The statistics of this measurement exceeds by a factor 3.5 those of the
previous measurement of Ref.~\cite{id}.
A few events with anomalously low yields are present, 
among which six populate the nuclear recoil band between 20 and 200 keV. 
This results in a measured rejection factor of $(3\pm1) \times 10^{-5}$.
The same rejection factor is measured with both analyses, 
and it does not depend on the tightness
of the $\chi^2$ cut (described in Section~\ref{selection}) 
used to increase the efficiency to remove mis-measured pile-up events.

The origin of the six events with anomalous charge yield is still under investigation. 
Simulations using the worst resolution of any detector show that no more
than 0.1 $\gamma$-ray events from the whole $^{133}$Ba calibration 
data should leak inside the NR band
because of Gaussian fluctuations on the heat and ionization measurements.
The anomalous events are not associated to a specific time period 
or a specific detector. 
In particular, none of the six events appearing in the nuclear recoil band 
are associated with bolometers which have a defective readout of an 
ionization channel. 
Since the gamma rejection is as good for these detectors as for the others, 
all ten detectors are used for the present WIMP search. 
Note that, based on these gamma calibrations, this includes one detector 
where the readouts of both a veto and a guard electrode are defective. 

Fig.~\ref{fig:gamma} also shows the line corresponding to an ionization yield 
of  3.72$\sigma$ below unity, for the average detector resolutions,
corresponding to the rejection of 99.99\% of all $\gamma$-ray 
interactions assuming a gaussian distribution of ionization yield.
From detector to detector, and depending on the instantaneous experimental
resolutions, this line intersects the NR band at recoil energies ranging from 10 to 20 keV, 
except for one detector where this value can reach 24 keV.
The large fluctuations of the $\gamma$-ray rejection performances below 20 keV
hinders the comparison of the results of the two analyses at these energies.
As in Ref.~\cite{idfirst}, the analysis threshold for the WIMP search 
is set to this value.

\section{WIMP search data}

\subsection{Data selection and exposure}
\label{selection}

The initial WIMP search data set comprises a total of 325 days for the ten detectors, 
plus 92 days for two detectors in the 2008 run.
In order to ensure a reliable background rejection, the data taking periods are selected 
hour-by-hour on the basis of the averages of heat and ionization baseline resolutions.
It is required that the hourly average of the fiducial ionization measurement 
is less than 2.0 keV FWHM,
and the corresponding resolution on both heat and guard ionization signals 
is less than 2.5 keV FWHM.
This selection results in a reduction of 17\% of the total exposure.
The additional loss due to acquisition dead-time is 3.2\%, 
a significant part of it being due to an imperfection in the trigger 
algorithm in the first months of data taking. 
To ensure a correct reconstruction quality, and to reject pile-up events,
noisy periods lasting less than one hour and events due to interactions inside the
Ge-NTD thermistors, cuts were applied on the $\chi^2$ of the pulse shape fits relative to known templates from calibrations, 
as well as on the reconstructed pulse position within the recorded ionization sample.
The efficiency of these cuts is measured using the gamma background
in WIMP search data, 
in particular the cosmogenic doublet at 9.0 and 10.4 keV which produces pulses 
with amplitudes similar to WIMP signals. 
The overall $\chi^2$ efficiency loss is $2.4\%$, 
while the ionization timing cut efficiency loss is $0.3\%$.

For the WIMP search, coincident events between two bolometers or with a trigger
in the muon veto within an appropriate time window are rejected, resulting
in a negligible deadtime.
  
The final exposure of 427~\kgd is calculated from the selected live time, 
corrected for the $\chi^2$ cut efficiency
and the effective fiducial mass value of 160~g. 
WIMP candidates are then selected in the 1.64$\sigma$ nuclear recoil band as described in Section 3.2. 
This 90\% efficiency results in an effective exposure of 384~\kgdf.

The efficiency as a function of measured recoil energy for WIMP interactions depends on
{\em i)} the level of the online heat trigger, {\em ii)} the offline
reconstruction cuts requiring that the heat (ionization) signals
be three (two) times above the measured value of
the instantaneous baseline FWHM resolution on the corresponding quantity,
and {\em iii)} the cut to reject electron recoils, requiring that
the normalized ionization yield be more than 3.72$\sigma$ 
below unity (corresponding to the rejection of 99.99\%
of all $\gamma$-ray interactions assuming a Gaussian
dispersion of the ionization yield).
This third factor has the strongest influence on the efficiency function. 
Overall, the relative acceptance for nuclear recoils is 98.3\% at 20 keV
and above 99.9\% at 23 keV and higher energy.

\subsection{Observed distributions}

\begin{figure}
\begin{center} \includegraphics[width=12cm]{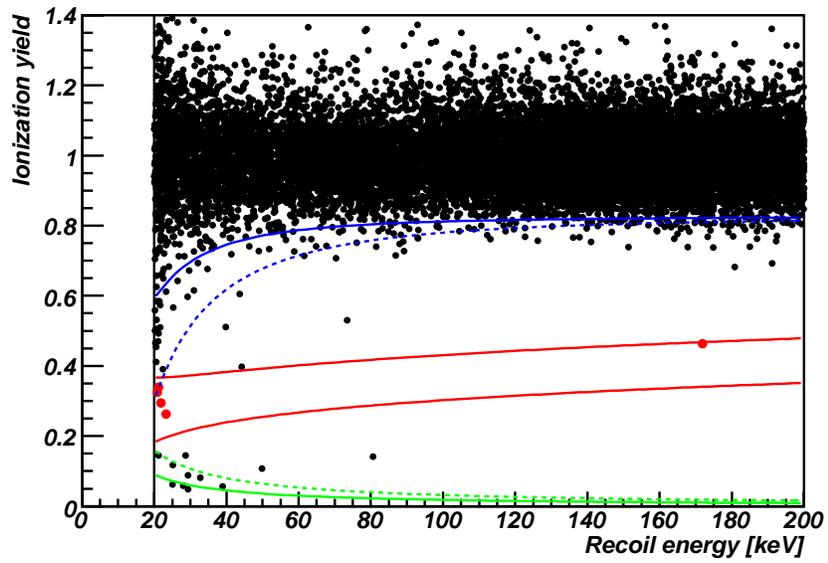} \end{center}
\caption{Ionization yield vs recoil energy of fiducial events recorded
by EDELWEISS-II in an exposure of 427~\kgdf.
The WIMP search region is defined by recoil energies between 20 and 200~keV, and an ionization yield 
inside the 90\% acceptance band (full red lines, corresponding to an effective exposure of 384 \kgdf). WIMP candidates are highlighted in red. The average (resp. worst) one-sided 99.99\% rejection limits for electron recoils are represented with a continuous (resp. dashed) blue line. The average (resp. worst) ionization thresholds are represented with a continuous (resp. dashed) green line.
\label{fig:qplot}}
\end{figure}

Fig.~\ref{fig:qplot} shows the scatterplot of ionization yield as a 
function of recoil energy for the WIMP search data over all detectors. 
The full green and blue lines represent the average ionization threshold of 2.0 keV
and 99.99\% electron-recoil rejection limit, respectively.
The dashed lines represent the corresponding worst values for any detector at any given time.
The online and offline heat thresholds have no significant impact 
on the data above 20~keV.
The red band represents the average nuclear recoil band for the ten detectors.

In the recoil energy range from 20 to 200 keV, a total of 
1.8$\times$10$^{4}$ fiducial events  are identified. 
The rate in the energy range from 20 to 50 keV is 
0.14 evts/keV/kg/day.
The bulk of the events have ionization yields within the region where
99.99\% of the electron recoils are expected, assuming a Gaussian
dispersion.
However, three populations are not consistent with electron recoils.
\begin{enumerate}
\item Eleven events lie well below the nuclear recoil band with recoil 
energies between 20 and 80 keV.
Interestingly, the four most energetic events are associated with a detector where a strong accidental
$^{210}$Po contamination was found on the detector holder~\cite{luca}.
This suggests that some of these anomalous events far from the ionization thresholds
could be due to a reduced rejection rate for degraded low-energy alphas relative to
 low-energy electrons. This population is still under investigation.
\item Four events between 40 and 80 keV lie above the NR band and 
have ionization yields of less than 0.65.
With the previous detector design without interleaved electrodes, these intermediate-yield 
events occurred at a more copious rate (by a factor $\sim$100~\cite{edw,silvia}) and
were associated with incomplete charge collection near the detector surface.
This type of event is largely suppressed in interleaved-electrode detectors~\cite{id}, 
resulting in the large gain in sensitivity relative to Ref.~\cite{edw}. 
However, the observed number of events exceeds
the prediction of $<1.5$ at 90\%CL derived from beta and gamma calibrations.
\item Five events appear in the nuclear recoil region. Four are between 20.8 and 23.2 keV, 
and one at 172 keV. One of them had already been reported in Ref.~\cite{idfirst}.
All five will be retained as nuclear recoil candidates in the WIMP search.
\end{enumerate}

Both analysis chains yield consistent results for these three populations.
In particular, the nuclear recoil candidate at 172 keV is also observed,
as well as four others at energies between 20.2 and 22.5 keV, although
the identity of the candidates differs as one event migrates
above the 20 keV threshold while another migrates below it
due to sub-keV fluctuations in energy reconstruction.

\subsection{Background estimates}
\label{section-bkg}

We now investigate the potential sources of background in the WIMP search region, 
using both calibration data, simulations and the measured backgrounds outside 
the nuclear recoil band. 
Three potential sources are considered: $\gamma$-rays, surface events and
neutron scattering.

The two main sources of $\gamma$-rays are the continuous background between
20 and 200 keV (1.8$\times$10$^4$ in the WIMP search data) and the
cosmogenic activation doublet at 10 keV (1$\times$10$^4$ events).
As discussed in the previous section, gaussian fluctuations of the
ionization and heat measurement cannot account for the presence
of events inside the nuclear recoil band above 20 keV.
In particular, assuming that the four observed events below 24~keV are due to
a 10.4 keV $\gamma$-ray requires fluctuations by 7 to 12 $\sigma$ 
on the fiducial ionization signals, depending on the event.
Non-gaussian fluctuations may be more important, for example those associated to events involving
an interaction near the surface of the guard region where the charge is not well collected.
But such fluctuations are difficult to predict with precision in a model-independent way.
However, an empirical estimation can be obtained
using the results of the $^{133}$Ba calibration, where a
a background of 3$\times$10$^{-5}$ NR candidates
per fiducial photon was observed.
As the spectrum between 20 and 200 keV is very similar
in $^{133}$Ba calibration and WIMP search runs,
it can be expected that the same process would proportionally yield
a background of less than 0.9 events in the NR band at 90\%CL.

The predicted number of unrejected surface events is estimated by multiplying
the number of observed low-ionization yield events before the rejection of
surface events ($\sim 5000$) by the upper limit on
the rejection rate measured in Ref.~\cite{id} (6$\times$10$^{-5}$ at 90\%CL).
This results in 0.3 events.
A deficient suppression of events due to surface $\beta$ contaminants is thus
an unlikely explanation for the events observed in the nuclear recoil band.
As another source of surface events, alpha radioactivity is 
estimated to generate a negligible leakage, 
from calibration measurements with alpha sources.

The muon veto efficiency was measured using two different methods,
one from internal coincidences within the veto, and the other using bolometer-veto
coincidences. 
The measured efficiency to veto a muon entering the cryostat
is compatible with 100\%, being larger than 92.8\% at 90\%CL.
The observation of 260 coincidences between the bolometers and the muon veto 
before any fiducial, energy or ionization yield cuts
corresponds to an average rate of muon-induced events of 0.17 $\pm$ 0.01 events per \kgdf.
Of these, 0.008 $\pm$ 0.004 events per \kgdf~appear as single events in the NR band above 20 keV.
Scaling this number to the exposure of the WIMP search data
and considering the lower limit on the muon veto efficiency measurement, this
 corresponds to an expected background of less than 0.4 events at 90\%CL in the present WIMP search. 
  
The estimate from Ref.~\cite{idfirst} for the contribution of neutrons from 
radioactive decays in the rock and concrete surrounding the experiment 
and the lead shield has been improved with more reliable GEANT4 simulations.
The simulation of the effect of the polyethylene shield on an external neutron 
flux was tested by comparison with data recorded with a strong neutron source (10$^5$ s$^{-1}$) 
positioned at different locations around the experiment outside the shields.
Following this work,
the upper limit on the number of nuclear recoil events due to the
flux of neutrons going through the polyethylene shield is 0.11.
The contribution from neutron sources inside the polyethylene shield
has been revised relative to Ref.~\cite{idfirst} following  
better measurements of the U/Th contents of some relevant materials,
and the study of additional sources.
The summed upper limit of the contributions from the contamination of the lead 
and polyethylene shields and their steel supports, as well as the copper
cryostat itself, is 0.21 events.
A potentially more important neutron source has been identified as
the connectors and cables located inside
the cryostat, which could induce up to 1.1 nuclear recoil events.

Summing all the 90\%CL upper limits from the different sources, 
we arrive at an estimated background of less than 3.0 events in 384~\kgdf. 
The Poisson probability to fluctuate from 3 to 5 events or more is 18\%. 
Interpreted as a central value, the background estimate indicates no evidence for WIMP events. 
However, in terms of understanding the nature of the observed background, the observation
of 5 events indicates that the well quantified part of our background model, 
corresponding to at most 3 events, fails to explain the data.
Given the small number of observed events, the 
data distributions in energy and ionization yield do not
help confirm or infirm the validity of part or all of the background model.
The statistics of the additional sample of events where more than one detector 
triggered in coincidence is not sufficient to yield useful informations 
on the nature of the background.
As most of the NR candidates appear close to  20 keV, 
the sample of NR events just below this analysis threshold was inspected.
This study is complicated by the appearance of significant systematic effects due 
to event calibration and reconstruction at 15 keV, as shown by the comparisons between 
the main analysis and the cross-check analysis.
After correcting for the inefficiency due to the gamma rejection cut, 
the number of events in the NR band between 15 and 20~keV varies 
between 10 and 20 per 384~\kgd depending on the analysis.
However a potential source for this population has been identified.
It consist of events where the charge is shared between the fiducial and
guard electrodes and collection on the latter electrode is suppressed at 
the level such that the guard signal passes the fiducial cuts.
The gamma-ray calibrations of events in the guard region indicate that
such important charge losses in that volume are not uncommon.
Simulations using as input the experimental ionization yield of guard events
and typical probabilities of charge sharing between the fiducial and guard
electrodes indicate that this mechanism can produce events inside the NR band 
at low energy at a rate comparable to what is observed between 15 and 20 keV. 
Unfortunately, the simulations fail to provide reliable quantitative estimates of 
this population, as they depend critically on details of the spatial distribution
of events and of the charge collection model in the guard region 
that are poorly constrained by experimental data.
It cannot be excluded that this population extends above 20 keV,
bringing closer the background estimate to the observed yield.

The background studies are being pursued as they provide useful informations 
to prepare the reduction of backgrounds necessary for the next phase of the experiment. 
These considerations do not affect the search results discussed below, 
based on the conservative assumptions that it cannot be excluded that part or all of the observed 
events in the NR band are due to WIMP-induced recoils.
Given the low statistics of this sample of events and the
uncertainties in the background prediction, the present WIMP search is performed 
by considering all five single events as valid nuclear recoil candidates.

\section{Limit on spin-independent cross-sections for elastic and inelastic models}

For both elastic and inelastic processes, upper limits on WIMP-nucleon cross-sections are 
calculated using the optimal interval method~\cite{yellin} and the prescriptions of Ref.~\cite{lewin}.
The halo model parameters are a local density of WIMP of 0.3~GeV/c$^2$, 
a Maxwellian velocity distribution with a {\em rms} velocity of 270~km/s, 
an average Earth velocity of 235~km/s, and a galactic escape velocity 544~km/s~\cite{rave}.
This later value has been chosen as the value of 650 km/s used previously overestimates the sensitivity of inelastic searches, while having a $<$0.5\% effect in the elastic case.
The calculation includes the effect of the finite recoil energy resolution of 
our detectors, which is on average 1.5~keV.

\subsection{Elastic cross-section}

\begin{figure}
\begin{center} \includegraphics[angle=90,width=12cm]{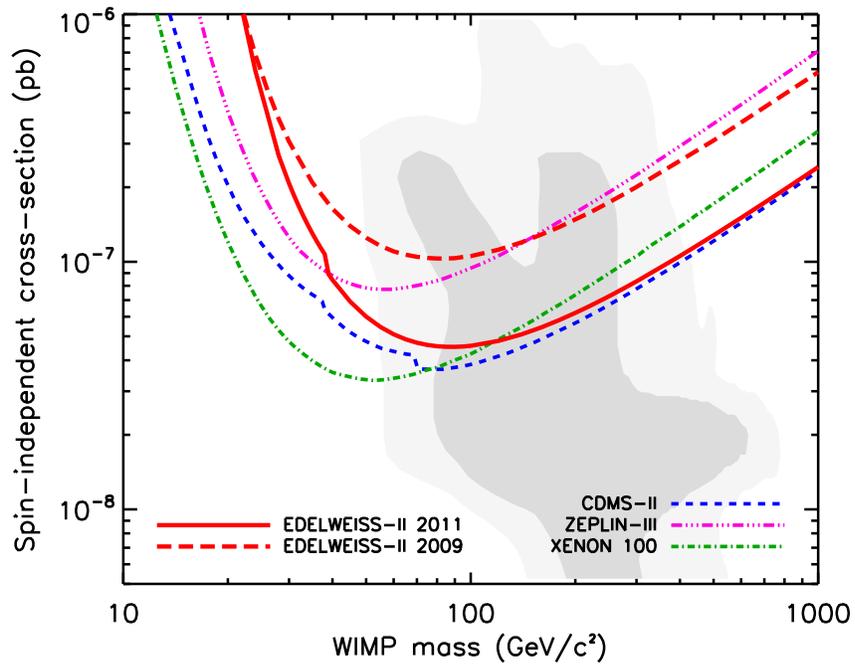} \end{center}
\caption{Limits on the cross-section for spin-independent scattering
of WIMPs on the nucleon as a function of WIMP mass, derived from the present work,
together with the limits from CDMS~\cite{cdms}, ZEPLIN~\cite{zeplin} and XENON100~\cite{xenon100}. The shaded area correspond to the 68\% and 95\% probability regions of the cMSSM scan from Ref.~\cite{theo}.
\label{fig:limits}}
\end{figure}

The 90\%CL upper limits on the WIMP-nucleon spin-independent cross-section 
derived from the present data are shown as a function of the WIMP mass in Fig.~\ref{fig:limits}.  A cross-section of $4.4\times 10^{-8}$~pb is excluded at 90\%CL for a WIMP mass of 85~GeV. 

This limit is more than twice as stringent than that obtained after the first six months~\cite{idfirst}. 
At higher WIMP masses, the improvement approaches the value of 2.7 that corresponds
to the increase of exposure, and the sensitivity becomes comparable to that of CDMS~\cite{cdms}.
It is better than that reported by XENON100~\cite{xenon100} above 110~GeV/c$^2$.
For lower WIMP masses, the Ge-ID sensitivity is degraded due to the presence of 
four events between 20.8 and 23.2~keV. 
The limits are not affected by the presence of the nuclear recoil candidate at 172~keV.

\subsection{Inelastic cross-section}

The inelastic dark matter scenario has been proposed to reconcile the 
dark matter modulation signal claimed by DAMA/LIBRA~\cite{dama} and the null results 
in all other direct detection experiments \cite{smithweiner}.
In this scenario, the WIMP-nucleus scattering would occur through a transition 
of the WIMP to an excited state $\delta\sim$100 keV heavier than the previous one. 
The elastic scattering would be highly suppressed.

Compared to the elastic case, the WIMP can
scatter off a nucleus only if its kinetic energy is sufficient to
induce the inelastic excitation.
The minimum WIMP velocity that can induce an observable recoil energy $E_R$ is:
$$ v_{\rm min}=\frac{1}{c^2}\sqrt{\frac{1}{2 M_N E_R}}\left( \frac{M_N E_R}{\mu} + \delta \right) $$
where $M_N$ is the mass of the target nucleus and $\mu$ is the reduced mass 
of the WIMP-target nucleus system.  
Therefore only the high end of the WIMP halo velocity distribution contributes to the signal: 
the event rate is globally reduced and suppressed at low recoil energies, 
and the modulation signal is enhanced.
The event rate is computed using the analytical solution of Ref.~\cite{savage}
and the same halo parameters as in the elastic case.

Mass splitting values ranging from $\delta =90$ to $\delta=140$ keV are 
particularly interesting since  CDMS~\cite{arrenberg2010} and 
XENON~\cite{xenon_idm} do not exclude all the DAMA-allowed region in that case.
Fig.~\ref{fig: edw vs cdms, delta120} shows the limit obtained for a 
mass splitting $\delta=120$~keV. 
Our limit excludes the DAMA region above $\sim$90 GeV/c$^2$, 
improving by $\sim$10 GeV/c$^2$ the CDMS limit of Ref.~\cite{arrenberg2010}.
For WIMP masses larger than $\sim$200 GeV/c$^2$, EDELWEISS excludes cross-section 
values that are half of those excluded by CDMS.
This is due,
to a large part, to the absence of WIMP candidates in the energy range 
between 23.2~keV and 172~keV, 
whereas CDMS observes three events in that same range.

\section{Conclusion}

The EDELWEISS-II collaboration has performed a direct search for
WIMP dark matter using ten 400-g heat-and-ionization cryogenic
detectors equipped with interleaved electrodes (Ge-ID) for the rejection
of near-surface events. 
A total effective exposure of 384~\kgd has been obtained in 
fourteen months of operation in 2009 and 2010 at the 
Laboratoire Souterrain de Modane and with
additional data from earlier runs with two  detectors in 2008.

This enlarged data set with respect to~\cite{idfirst} is used to set limits on 
the cross-section of spin-independent interactions of  WIMPs and nucleons. 
Five nuclear recoil candidates are observed above a recoil energy 20 keV.
A value of the elastic spin-independent cross-section of $4.4\times10^{-8}$~pb 
is excluded at 90\%CL for a WIMP mass of 85~GeV.
At larger masses, the sensitivity becomes comparable to the one obtained
by the experiment CDMS.
It confirms their results in an independent experiment using germanium 
detectors with a different design.
The Ge-ID data also improve the constraints on models of inelastic dark matter with
a WIMP mass splitting around 120~keV. 

The fact that these significant results have been achieved
with the very first array of these new detectors
attests the competitiveness of the Ge-ID technology in direct WIMP searches.
The background studies performed in the present work have fostered progresses 
on two experimental issues.
First, the measurement and control of the presence of U/Th traces inside the cryostat 
is being improved, and an inner polyethylene shielding will be installed. 
Second, a new detector design with a smaller fraction of non-fiducial volume is now used, 
in order to reduce the role of mis-measured ionization signals 
from this region, as discussed in Section~\ref{section-bkg}. 
With these improvements and the coming increase by an order of magnitude of
the mass of the array, the goal is to soon probe the physically interesting 
range of spin-independent  WIMP-nucleon cross-sections between 
10$^{-8}$ and 10$^{-9}$~pb.

\section{Acknowledgments}
The help of the technical staff of the Laboratoire 
Souterrain de Modane and the participant laboratories is 
gratefully acknowledged. This project is supported in part by the
Agence Nationale pour la Recherche under contract ANR-06-BLAN-0376-01, 
by the Russian Foundation for Basic Research
and by the Science and Technology Facilities Council, UK.

\newpage

\begin{figure}
\begin{center}
\includegraphics[angle=90,width=12cm]{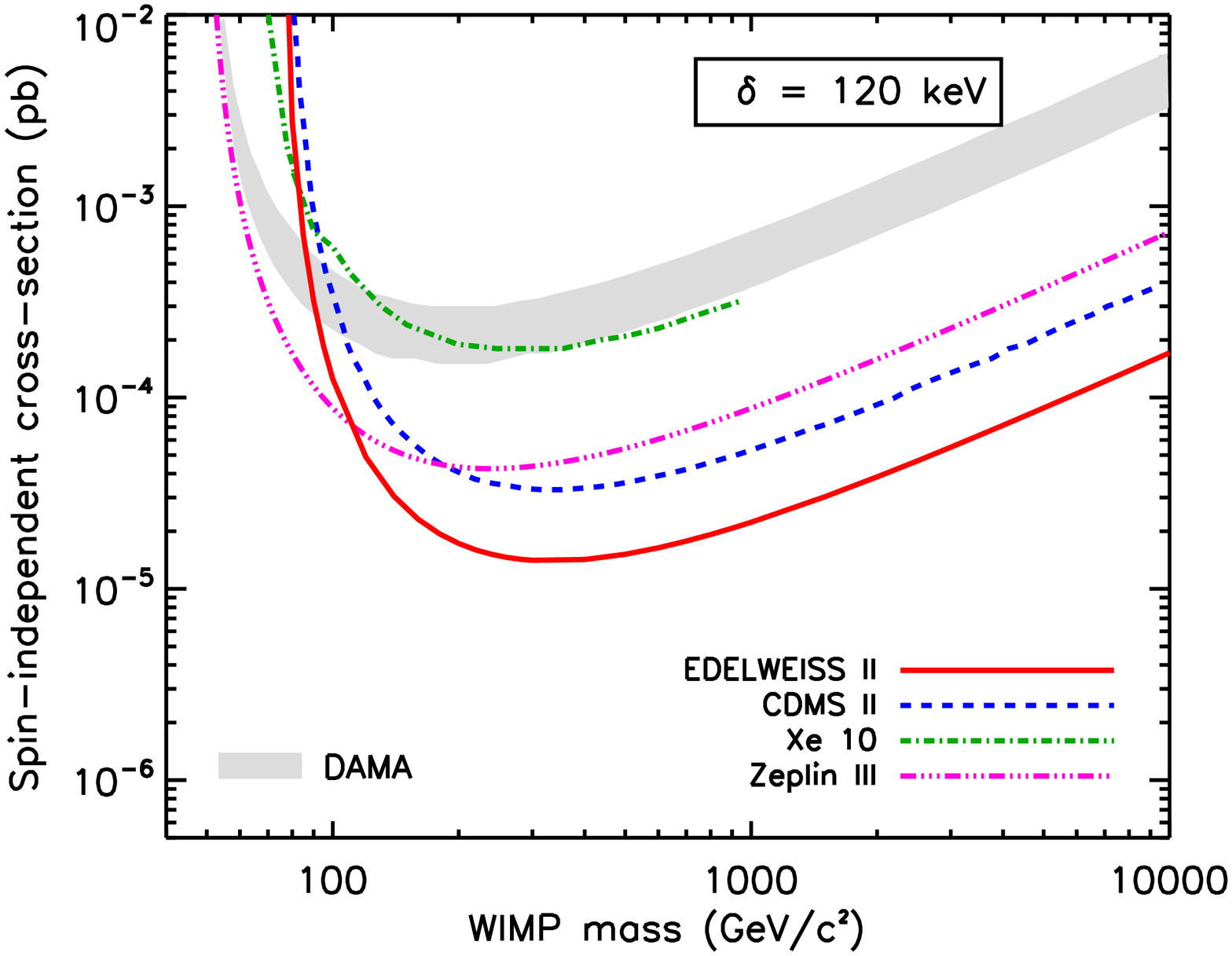}
\caption{Inelastic WIMP-nucleon cross-section limits at 90\%CL as a function of WIMP mass, for a mass splitting value $\delta
=120$ keV. Also shown are the limits from XENON10~\cite{xenon_idm} (with a conservative $L_{\rm eff}$), ZEPLIN-III~\cite{zeplin_idm} and CDMS~\cite{arrenberg2010} (from a dedicated analysis). The 95\% allowed DAMA contour, as estimated in~\cite{arrenberg2010} from~\cite{dama}, is shown in light gray.}
\label{fig: edw vs cdms, delta120}
\end{center}
\end{figure}


\begin{thebibliography}{10}

\bibitem{cosmo} E. Komatsu et al., Astrophys. J. Suppl. 192 (2011) 18.
\bibitem{wimp} G. Bertone, D. Hooper and J. Silk, Phys. Rep. 405 (2005) 279.
\bibitem{goodman} M. W. Goodman and E. Witten, Phys. Rev. D 31 (1985) 3059.
\bibitem{rev} G. Jungman, M. Kamionkowski, and K. Griest, Phys. Rep. 267 (1996) 195; \\
R. J. Gaitskell, Annual Rev. Nucl. and Part. Sci. 54 (2004) 315.
\bibitem{brink} P.L. Brink, et al., Nucl. Instrum. Meth. A 559 (2006) 4148.
\bibitem{broni} A. Broniatowski, et al., J. Low Temp. Phys. 151 (2008) 830.
\bibitem{id} A. Broniatowski et al., Phys. Lett. B 681 (2009) 305.
\bibitem{idfirst} E. Armengaud et al., Phys. Lett. B 687 (2010) 294.
\bibitem{smithweiner} D. Smith and N. Weiner, Phys. Rev. D 64 (2001) 043502.
\bibitem{edwsetup} EDELWEISS collaboration, Paper in preparation.
\bibitem{lemrani} R. Lemrani et al., J. Phys. Conf. Ser. 39 145 (2006) 145.
\bibitem{martineau} O. Martineau et al., Nucl. Instrum. Meth. A 530 (2004) 426.
\bibitem{luca} L. Pattavina, Ph.D. thesis, Universita Degli Studi di Milano-Bicocca (Milano, Italy) 
and Universit\'e Claude Bernard Lyon 1 (Lyon, France) (2011).
\bibitem{edw} V. Sanglard et al., Phys. Rev. D 71 (2005)122002.
\bibitem{silvia} Silvia Scorza, Ph.D. thesis, Universit\'e Claude Bernard Lyon 1 (Lyon, France) (2009).
\bibitem{yellin} S. Yellin, Phys. Rev. D 66 (2002) 032005.
\bibitem{lewin} J.D. Lewin and P.F. Smith, Astropart. Phys. 6 (1996) 87.
\bibitem{rave} M. C. Smith et al., Mon. Not. Roy. Astron. Soc. 379 (2007) 755-772.
\bibitem{cdms} Z. Ahmed et al., Phys. Rev. Lett. 102 (2009) 011301;\\
Z. Ahmed et al., Science 327 (2010) 1619-1621.
\bibitem{xenon100} E. Aprile et al., Phys.Rev.Lett.105 (2010) 131302.
\bibitem{dama}R. Bernabei et al., Eur. Phys. J. C 67, 39 (2010).
\bibitem{savage} C. Savage et al., Phys. Rev. D 74 (2006) 043531.
\bibitem{arrenberg2010} S. Arrenberg et al., Phys. Rev. D 83 (2011) 112002.
\bibitem{xenon_idm} J. Angle et al., Phys. Rev. D 80 (2009) 115005.
\bibitem{theo} L. Roszkowski, R. Ruiz de Autri and R. Trotta, JHEP 07 (2007) 075.
\bibitem{zeplin} V.N. Lebedenko et al., Phys. Rev. D 80 (2009) 052010.
\bibitem{zeplin_idm} D.Y. Akimov et al., Phys. Lett. B 692 (2010) 180-183.

\end{thebibliography}
\end{document}